\documentclass{article}

\usepackage[dblblindworkshop, final]{neurips_2026}
\workshoptitle{AI4Mat 2026}
\usepackage{amsmath}
\usepackage{amssymb}
\usepackage{float}
\usepackage{graphicx}

\usepackage[utf8]{inputenc} % allow utf-8 input
\usepackage[T1]{fontenc}    % use 8-bit T1 fonts
\usepackage{hyperref}       % hyperlinks
\usepackage{url}            % simple URL typesetting
\usepackage{booktabs}       % professional-quality tables
\usepackage{amsfonts}       % blackboard math symbols
\usepackage{nicefrac}       % compact symbols for 1/2, etc.
\usepackage{microtype}      % microtypography
\usepackage{xcolor}         % colors

\newcommand{\pT}{p_{\mathrm{T}}}
\newcommand{\dpT}{\Delta \pT}
\newcommand{\dKU}{\Delta_{\mathrm{KU}}}

\title{What Do LLMs Use to Judge Synthesis Recipes? An Experimental Analysis on 240 Solid-State Syntheses}

\author{
\textbf{Derek van Tilborg, 
Laura-Bianca Pasca,
Harry Sevi, 
Kyle Reeves,
Thomas Loux}, \\
\textbf{Aikaterini Flessa Savvidou,
Varincka Jean-Gilles,
Marilou Carouge,} \\
\textbf{Joonatan E.\,M.\ Laulainen} \\
[4pt]
Altrove, 25 Rue de Ponthieu, Paris, France \\
\texttt{joonatan.laulainen@altrove.ai} \\
}

\begin{document}

\maketitle

\begin{abstract}
Inorganic materials discovery is increasingly limited by synthesis rather than by material design. Large language models (LLMs) are being adopted to propose synthesis conditions with little experimental evidence of their capabilities and biases. Here, we carry out an experimental synthesis benchmark of 240 experiments across eight known and unknown inorganic oxides and subsequently evaluate an LLM as a zero-shot ranker over fixed banks of solid-state synthesis recipes. For each pair of recipes for the same oxide target, \texttt{GPT-5.6-Sol} selects the recipe expected to yield higher target-phase purity. We aggregate these pairwise preferences using a regularized Bradley–Terry model to obtain a ranked list of synthesis recipes for a given target material. By comparison with the actual purity, obtained through X-ray diffraction, we measure \texttt{GPT-5.6-Sol}’s performance as a zero-shot ranker (0.72 pairwise accuracy). By selective removal of the information exposed to the model during this ranking, we conclude that ranking is driven predominantly by thermal features, with reaction chemistry modulating the effect. Ranking accuracy tracks the purity–temperature correlation by target (Spearman $\rho=0.91$, $p = 0.002$), and on the six thermally-driven targets withholding all chemical identity hardly impacts performance (0.68) compared to withholding temperature (0.56). Ranking performance was higher on known than unknown targets, but the group-level difference was not statistically conclusive. The methodological choices highlight the need for more experimentally validated synthesis benchmarks and position LLMs as a thermally informed prior.
\end{abstract}

\section{Introduction}

Generative models and high-throughput computational screening now propose candidate inorganic compounds far faster than laboratories can attempt to experimentally validate them \citep{Merchant2023, Zeni2025}. As a result, the bottleneck in materials discovery has shifted from proposing compounds to synthesizing them \citep{park2026closing}. This imbalance has motivated automated laboratories which, in turn, require solid-state synthesis algorithms \citep{Szymanski2023, Szymanski2023_ARROWS3}. However, solid-state synthesis lacks robust simulation models. Instead, synthesis planning has become data-driven, with attention turning increasingly towards large language models (LLMs). What remains lacking is a structured experimental benchmark on which these models can be tested, particularly one containing materials that have not previously been synthesized. In this work, we construct such a benchmark and use it to characterize LLM performance and bias during synthesis planning.

Automated synthesis planning has primarily focused on predicting firing temperatures and precursors for a given target. Corpora text-mined from the literature \citep{Kononova2019, Lee2025} have enabled models that recover processing conditions to approximately 140\,$^\circ$C mean absolute error \citep{Huo2022} and precursor sets with at least 82\% success when five candidate sets are proposed \citep{He_2023}, while a separate line of work derives routes from thermochemistry rather than from text \citep{Aykol_2021, McDermott2021}. Both have been deployed in autonomous laboratories \citep{Szymanski2023}, although a re-examination of one such laboratory's products identified recurring shortfalls in phase identification \citep{Leeman2024}. LLMs have since been applied to the same tasks \citep{kim2024large, Prein2025SyntMTE} and have even been given access to laboratory hardware \citep{Boiko2023, Bran2024}. A body of work now tries to improve LLM-driven synthesis by injecting additional priors into them, with varying levels of success (Appendix~\ref{app:priors}). Approaches fusing Bayesian Optimization and LLM protocols have recently been proposed as a balance between LLM general chemistry knowledge and GP data-driven learning \citep{Rankovic2026}.

Reported performance on these tasks is high, but its interpretation depends strongly on how the evaluation was constructed. For example, synthesizability classification reaches 98.6\% accuracy against a negative class of 80,000 computationally selected structures rather than observed failures \citep{Song2025}. Precursor prediction was scored by exact match against the single recipe on record, which its own authors describe as a lower bound that pretraining leakage may inflate \citep{Prein2025SyntMTE}. However, expert and laboratory evaluations reach a different result \citep{Yanguas_Gil_2025, Prein2025Pipeline}. These discrepancies originate in the data itself: literature-derived recipes lack clear comparatives, are frequently only partially described, rarely report purity, and almost never report failures, making it is difficult to establish that one recipe is better than another. Current benchmarks therefore may mostly measure re-discovery, in the sense that the material has been made before and a route towards it is described somewhere in the training corpus. LLM performance is known to be artificially inflated in such cases in non-materials contexts \citep{sainz2023nlp, deng2024investigating}. This is directly in conflict with the goal of materials discovery, where automated synthesis is primarily concerned with materials that have not previously been synthesized.

In this work, we present an experimental benchmark for solid-state synthesis and use it to evaluate an LLM as a pairwise discriminator and ranker of synthesis recipes. To our knowledge, it is the first synthesis benchmark assembled from measured outcomes on a condition grid shared across targets rather than from published recipes, and the first evaluation of an LLM as a ranker over such a bank. In a pairwise comparison, we present the LLM with two recorded recipes for the same target and ask which of the two will produce more of the target phase. Every unordered pair within a target is judged in both presentation orders, and the resulting preferences are aggregated with an $\ell_2$-regularized Bradley--Terry model \citep{bradley1952rank} into a latent score for every recipe in the bank. These scores induce an ordering over candidates, whose utility we later test in a retrospective active-learning simulation using the pre-collected experimental outcomes. We use \texttt{GPT-5.6-Sol} \citep{OpenAI2026GPT56Sol} as the pairwise judge throughout.

The benchmark consists of experiments we conducted on each of eight oxide targets across 30 experimental profiles. We recorded a purity for 233 of the 240 attempts; 7 unsuccessful syntheses are present by construction. Four targets have published routes, whereas three have no prior experimental report of any kind and are not disclosed here, so no model can have encountered them during pretraining. One material is excluded from known/unknown comparison (see \S \ref{sub:setup}). Recipes are compared only within a target, so the task measures whether the model can recover the relative ordering of processing conditions rather than compare the intrinsic synthesizability of different materials. This design allowed us to reveal important features of LLM behavior in solid-state synthesis.

\section{Methods}
\subsection{Experimental Setup}
\label{sub:setup}
We construct a solid-state synthesis benchmark of eight known and unknown oxide materials, each synthesized across the same predetermined set of reaction conditions. To prevent leakage into the general LLM corpus we do not disclose the target materials, with one exception: Material 7 is Li$_7$La$_3$Zr$_2$O$_{12}$, disclosed so that the synthesis and characterization process can be inspected in full.

\begin{table}[t]
\centering
\small
\caption{\textbf{The eight benchmark targets and how the ranker performs on each.} 'Reports' counts prior experimental reports; 'Recipes' the conditions with a measured purity $\pT$, out of 30 attempted. The 'Ranker' block is \texttt{GPT-5.6-Sol}'s performance on the full prompt: pairwise accuracy ($\mathrm{Acc.}$), the Spearman correlation between its Bradley--Terry ranking and the true $\pT$ ranking ($\rho_{\mathrm{BT}}$), and the flip rate between the two presentation orders ($\mathrm{Flip}$). $\rho_T$ is the correlation between peak firing temperature and $\pT$ across each recipe for a given material; $\eta$ is the LLM's end-state ranking efficiency in the campaign of Section~\ref{sec:al}. M4 has an unknown ordered and a well known disordered phase we cannot reliably distinguish, so it is reported but excluded from every known-novel comparison. M7 is Li$_7$La$_3$Zr$_2$O$_{12}$. Known and unknown means are averaged over number of non-tied duels.} 
\label{tab:targets}
\begin{tabular}{llrrcrrrrr}
\toprule
& & & & & \multicolumn{3}{c}{Ranker} & & \\
\cmidrule(lr){6-8}
Material & Novelty & Reports & Recipes & $\pT$ range & Acc. & $\rho_{\mathrm{BT}}$ & Flip & $\rho_T$ & $\eta$ \\
\midrule
M1 & unknown  & 0    & 27 & 0.00--0.41 & 0.62 & $+0.17$ & 0.14 & $-0.55$ & 0.67 \\
M2 & unknown  & 0    & 30 & 0.00--0.27 & 0.45 & $-0.34$ & 0.09 & $+0.15$ & 0.41 \\
M3 & unknown  & 0    & 30 & 0.00--0.96 & 0.71 & $+0.54$ & 0.10 & $+0.46$ & 0.79 \\
M4 & excluded & --   & 30 & 0.00--0.92 & 0.73 & $+0.68$ & 0.18 & $+0.76$ & 0.80 \\
M5 & known    & 1--3 & 30 & 0.00--1.00 & 0.75 & $+0.63$ & 0.06 & $+0.61$ & 0.83 \\
M6 & known    & 1--3 & 30 & 0.00--0.92 & 0.86 & $+0.80$ & 0.05 & $+0.79$& 0.87 \\
M7 & known    & many & 27 & 0.00--0.95 & 0.66 & $+0.43$ & 0.07 & $+0.49$ & 0.70 \\
M8 & known    & many & 29 & 0.00--0.83 & 0.85 & $+0.83$ & 0.01 & $+0.85$ & 0.88 \\
\bottomrule
\end{tabular}
\end{table}

\begin{figure}[t]
  \centering
  \includegraphics[width=1.0\linewidth]{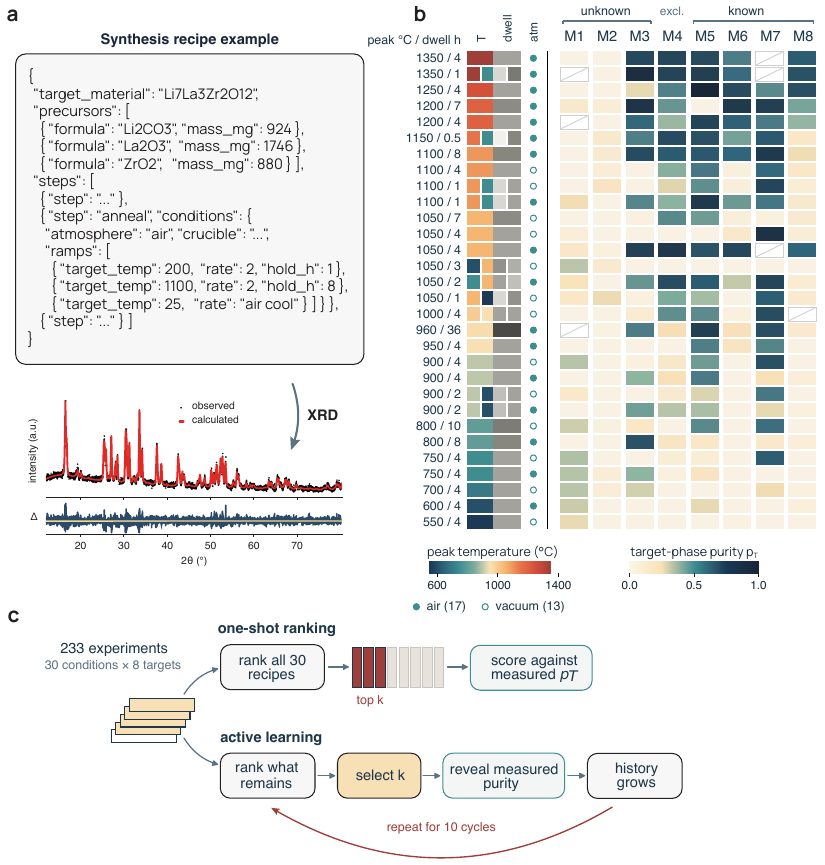}% use vector if possible
    \caption{\textbf{a.} Each experiment is a structured recipe whose product is measured by XRD, giving the target phase purity $\pT$. The profile is an example of a fit on Material 7, Li$_7$La$_3$Zr$_2$O$_{12}$. \textbf{b.} 30 firing conditions $\times$ 8 oxide targets, 233 measured (hatched = not measured). Strips give peak temperature, dwell and atmosphere; cell color is $\pT$. Targets are grouped by whether a published route exists; M4 is excluded. \textbf{c.} The two protocols: rank all 30 recipes once (one-shot ranking), or rank, run $k$, and repeat for 10 cycles (active learning).}
  \label{fig:method}
\end{figure}

\textbf{Materials Choice.} The targets, referred to as Material 1 to Material 8 (M1-M8), are listed in Table~\ref{tab:targets}, and their temperature profiles, atmospheres, and resulting purities are shown in Fig.~\ref{fig:method}. All eight are oxides, two ternary and six quaternary, with twelve distinct cations across the set. Precursors and their amounts were deliberately held fixed within a target, so that what varies across a target's bank are the firing conditions and little else.

\textbf{Material Novelty.} Materials 1, 2 and 3 have no prior experimental evidence of synthesis and were drawn from a bank of publicly unknown materials synthesized in our laboratory. Materials 5 and 6 are known but thinly documented, with one to three literature reports each, and Materials 7 (Li$_7$La$_3$Zr$_2$O$_{12}$) and 8 are well known. Material 4 has an unknown ordered phase and a well known disordered phase that we cannot reliably tell apart by X-ray diffraction; we collate the two for measurement and exclude it from every known-novel comparison, which leaves four known and three novel targets.

\textbf{Temperature Profile Choice.} We carried out the same 30 condition profiles on every target, chosen to spread widely across peak temperature, dwell time, number of holds and atmosphere. Purity depends strongly on where a recipe sits in that space (Fig.~\ref{fig:method}): some targets formed only at high temperature (M7 and M8), others across the range (M3-M6), and one only at low temperature (M1). Milling parameters were varied as well but had little effect on purity. The set is therefore thermally driven on average, a caveat that applies to all results below.

\textbf{Material Characterization and Benchmark Construction.} Every experiment was characterized by X-ray diffraction (XRD) on a Malvern Panalytical Aeris benchtop diffractometer, with spectra analyzed by an in-house algorithm and every fit checked by hand (Appendix~\ref{app:xrd}). Of the 240 attempts, 233 returned a measured purity; the remaining 7 samples fused to their crucibles, evaporated, or were lost (Fig.~\ref{fig:method}). The final benchmark is a single table of recipe and target phase purity $\pT$ (Rietveld-refined weight-\%).

\subsection{Pairwise Benchmark Measurement}

We evaluate the LLM as a ranker rather than as a generator of recipes or a predictor of purity: it is never asked to write a synthesis route and never asked to return a number. It is presented with two recorded recipes for a single target and asked which will reach higher target phase purity, and these preferences are aggregated into a ranking of that target's bank. We use \texttt{GPT-5.6-Sol} for the primary results and reproduce them with a smaller model in Appendix~\ref{app:models}.

\textbf{Pairwise Setup.} The benchmark comprises target materials $\mathcal{T}$ with $|\mathcal{T}| = 8$. Each target $t$ has a bank of recorded recipes $R_t$ with $n_t = |R_t| \in [26, 30]$ and $\sum_t n_t = 229$. A recipe is a sequence of process steps: milling programs, and furnace segments with ramp rates, hold temperatures, hold durations and atmosphere. The judge is a map
$$J:\ (x, x') \mapsto \{A, B, \varnothing\},$$
taking an ordered pair of recipe presentations and returning which of the first ($A$) or second ($B$) it expects to reach higher purity, with $\varnothing$ for an unparseable or failed response; the output format admits no other verdict. A presentation $x = \phi(r)$ renders recipe $r$ under a fixed template covering its thermal, precursor, milling and furnace fields. The judge sees no purity, phase analysis, or identifier. The template and an example prompt are provided in Appendix~\ref{app:prompt}.

\textbf{Pair Comparison Construction.} For each target we take the complete unordered pair set $P_t$, giving $\sum_t |P_t| = 3170$ pairs; pairs are never formed across targets, since purity is only comparable within a target. Every pair is judged twice, once in each presentation order, giving $\sum_t |D_t| = 6340$ duels per prompt condition. Both orders are run to cancel any preference for whichever recipe is presented first. Additionally, the rate of disagreement between them measures the consistency of the model's judgment on that target. Duels are identified by a condition-independent hash, so every condition is evaluated on exactly the same pairs (Appendix~\ref{app:duelid}).

\subsection{Scoring}

\textbf{Pairwise accuracy.} A duel is scored correct when the judge names the recipe with the higher measured purity. Per-target accuracy is the unweighted mean over that target's scorable duels, with chance at $1/2$. Both orderings of a pair count as one observation (\textit{i.e.}, a pair whose orders disagree contributes 0.5 rather than two trials). Duels with no parsable verdict are excluded, as are pairs of identical measured purity, for which no answer is correct.

\textbf{Purity gap and tie exclusion.} The purity gap of a pair is $\delta = |\,p(r) - p(r')\,|$. When $\delta = 0$ neither recipe is the purer one and the duel has no correct answer, so those pairs are excluded; every pair with $\delta > 0$ is scored. Of the 3{,}170 pairs, 375 are excluded on this basis, leaving 2{,}795 scored pairs per prompt condition. The gap determines the difficulty of a duel.

\textbf{Bradley--Terry score.} To convert the pairwise judgments into a single ranking, we fit a Bradley--Terry model in which each recipe $i$ is assigned a latent strength $\theta_i$, with $P(i \succ j) = \sigma(\theta_i - \theta_j)$ for the logistic function $\sigma$. Strengths are fit by minimizing the $\ell_2$-regularized negative log-likelihood over the observed outcomes, and ranking recipes by $\theta_i$ yields the model's prioritized ordering of the bank. The optimizer, regularization strength and identifiability constraint are given in Appendix~\ref{app:bt}.

\textbf{BEDROC score.} To evaluate early recognition in a campaign, which rewards methods that put successful recipes near the top of a prioritized list, we compute the Boltzmann-Enhanced Discrimination of Receiver Operating Characteristic (BEDROC) \citep{BEDROC_2007}. BEDROC applies an exponential weight $\alpha$ that penalizes successes found late, over the ranks of the $n$ active candidates ($\pT \ge 0.75 \times \pT^{\max}$ for that material) within a bank of $N$. We fix $\alpha = 20$, which concentrates roughly 80\% of the weight onto the top 8\% of the ranking. The full expression is given in Appendix~\ref{app:bedroc}.

\textbf{Ranking efficiency.} BEDROC depends on how many recipes were run in total, so it rewards depleting a bank as well as ordering it well. To isolate ordering quality we also report the ranking efficiency, which measures only whether the recipes a method acquired were acquired in a good order. Let $i = 1, \dots, m$ index those recipes in acquisition order and $p_{\mathrm{T},i}$ be the measured purity of the $i$-th, so that
$$\eta = \frac{\rho + 1}{2}, \qquad \rho = \rho_{\text{S}}\!\left(-i,\; p_{\mathrm{T},i}\right),$$
where $\rho_{\text{S}}$ is the Spearman rank correlation. The sign on $i$ makes earlier acquisition count as better. The map onto $[0, 1]$ normalizes each material against the best and worst possible orderings of the same recipes, so $\eta = 1$ is best-first, $0.5$ arbitrary and $0$ worst-first. Being a rank correlation over a material's own recipes, $\eta$ is invariant to how the budget was distributed across materials and requires no success threshold. We report $\eta$ averaged over a target material.

\subsection{Negative control}

We cannot control for the difficulty of synthesizing the different targets, since a model could perform well on a target purely because that target rewards common synthesis features. We therefore measure matched controls, in which the same recipe pairs are judged again with information removed, so that interpretable signals can be separated from per-target effects.

\textbf{Target and precursor masking.} We repeat the full pair comparison with the target identity and the precursor formulas masked out, so that the model is given no information about what is being made, only how it was processed.

\textbf{Temperature masking.} To mask thermal features, we remove every temperature field, including the derived thermal budget.

The comparison between the two controls constitutes the main result of Section~\ref{sec:pairwise}. Further controls, in which we mask the atmosphere, mask the recipe in its entirety, and add rather than remove information, are reported in Appendix~\ref{app:controls}.

\subsection{Active Learning Methods}
\label{sec:al-methods}

\textbf{Active learning campaign.} The pairwise task measures prediction accuracy, however, in a laboratory setting prioritizing high target purity recipes (and successfully synthesizing materials sooner) matters more than prediction accuracy. To mirror an experimental discovery campaign, we select the top 3 ranked recipes each round, reveal their measured purities, and repeat until the candidate bank is depleted, which takes 9 or 10 rounds depending on the target. The model knows a recipe's outcome only after the round in which it was selected. We compare the BT LLM ranker against a Bayesian optimization protocol using a Gaussian process (GP) surrogate, from which a log Expected Improvement acquisition function ranks them. The GP is refit on all data acquired so far at the start of each round, whereas the LLM is never retrained. We additionally include a random control and a thermal heuristic that ranks recipes by $1000 \times$ peak temperature plus total furnace hours.

Lacking data in the first round, the GP's opening batch is instead selected by a deterministic rule prioritizing variety across the accessible processing conditions. From the second round onwards it is fit on five recipe features: peak firing temperature, total dwell time, the presence of more than one firing step, the use of an air atmosphere at any point, and the thermal budget. The cold-start rule, kernel and hyperpriors follow \citet{chen_2026} and are described in full in Appendix~\ref{app:gp}.

\section{Pairwise Results}
\label{sec:pairwise}

\textbf{Ranking performance across the benchmark.} All eight targets were synthesized under some subset of the conditions, with M1 and M2 forming only as a minority phase, which caps their attainable purity. The pairwise ranking is 0.72, but varies considerably across the set (Table~\ref{tab:targets}): pairwise accuracy ranges from 0.86 on M6 to 0.45 on M2, the latter below random chance.

\begin{figure}[t]
  \centering
  \includegraphics[width=1.0\linewidth]{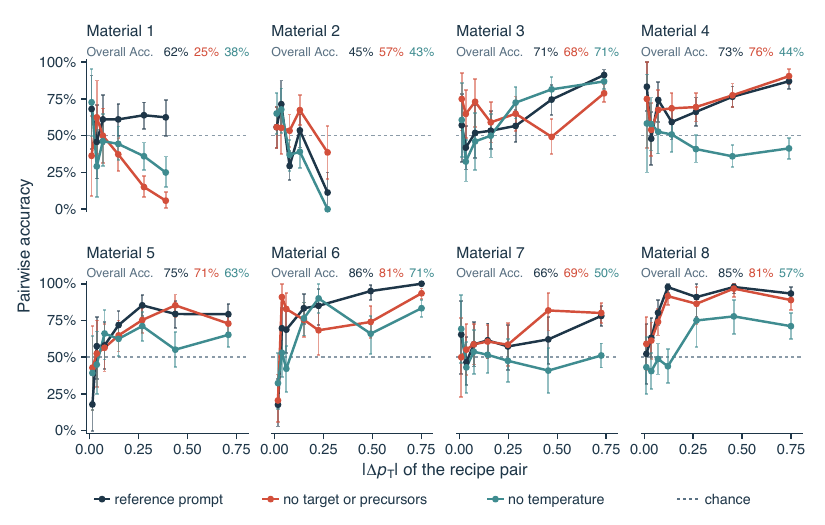}
    \caption{\textbf{Pairwise recipe ranking performance per target.} Pairwise accuracy across pair purity differences $|\dpT|$, for each of the eight targets. \texttt{GPT-5.6-Sol} ranking performance is shown for the reference prompt (black), without including the target formula and precursors (red), and without temperature information (teal). Overall pairwise accuracy (regardless of $|\dpT|$) is annotated for each material. Accuracy of a well-performing ranker is expected to rise for higher $|\dpT|$.}
  \label{fig:per_target_pairwise}
\end{figure}

\textbf{Accuracy by purity gap.} Since the difficulty of a duel is set largely by how far apart the two recipes are in measured purity, we investigated accuracy in relation to the purity gap $|\dpT|$ (Fig.~\ref{fig:per_target_pairwise}). This enables clearer comparison between materials with different difficulty of synthesis. Accuracy on targets M3--M8 increases monotonically; accuracy on M1 is monotonic and flat around 0.6, and remains about 0.5 for M2. Per-target accuracy is strongly rank-correlated with the correlation between purity and peak firing temperature $\rho_T$ ($\rho_S$ = +0.91, t(6) = 5.20, two-sided p = 0.002; exact permutation p = 0.005; n = 8), and more strongly still with the Bradley–Terry rank correlation $\rho_{BT}$ ($\rho_S$ = +0.95, exact p = 0.001). The pairwise accuracy converges to 0.5 at 0 purity difference, as expected (see per-class average in Fig. \ref{fig:ku_discrimination}). This suggests that the model may be primarily leveraging target-independent thermal features, rather than the target chemistry.

\textbf{Ranking without chemical information.} To further probe the effect of the target chemistry, we removed all chemical identity from the recipes and performed new duels, leaving only the firing conditions (Fig.~\ref{fig:per_target_pairwise}). If the model had a strong understanding of target chemistry, removing that information should degrade performance significantly. Overall pairwise accuracy only drops by $-0.04$ to $0.68$, with rank correlation falling by only $0.096 \pm 0.132$ ($p = 0.58$, two-sided exact sign-flip permutation test). The most significant pairwise accuracy is lost on M1 (loss of $-0.37$), where the temperature correlation $\rho_T$ is negative. In contrast, on the thermally dominated materials (M3--M8), the pairwise accuracy drops only from $0.76$ to $0.74$; implying that the high performance is almost entirely due to the underlying thermal profile rather than \texttt{GPT-5.6-Sol} resolving the chemistry. Removing the atmosphere, which is also relevant to the chemistry, drops the performance by $-0.04$ to $0.68$ while notably degrading performance on M3 (loss of $-0.29$, Table \ref{tab:per_target_accuracy}). The other thermally dominated targets lose an average of $0.02$ (M4 excluded). Both the effect on M1 and M3 reduce when switching to \texttt{GPT-5.6-Luna} (\ref{app:controls}). Performance on M2 and M7 increases when the chemistry is removed, implying that knowing it is occasionally unhelpful. Overall, this suggests the model's pairwise performance is dominated by overarching thermal heuristics, but that chemistry can modulate its performance (M1 and chemistry, M3 and atmosphere).

\textbf{Ranking without temperature.} Removing the temperature fields instead, including the derived thermal budget, has considerably more impact on performance (Fig.~\ref{fig:per_target_pairwise}). Rank correlation drops by $0.362 \pm 0.104$ ($p = 0.02$) and pairwise accuracy by $0.137 \pm 0.040$ ($p = 0.03$), both two-sided exact sign-flip permutation tests over the targets. Temperature is thus the channel through which most of the ranking runs, and removing it has a far greater effect than the chemistry.

The targets in this benchmark are largely thermally driven: the correlation between purity and peak firing temperature averages $+0.68$ across the known targets and $+0.02$ across the novel ones (Table~\ref{tab:targets}). Note that this follows conventional solid state synthesis principles \citep{stein1993turning}. The model prefers them as well, selecting the hotter recipe in 72\% of the comparisons in which the two differ in peak temperature (per-material breakdown of LLM agreement with the heuristic baseline is shown in Fig.~\ref{fig:SI-heuristic-agreement}).

\textbf{Known and novel targets.} To quantify the difference between re-discovery and discovery, we compared ranking performance across the known and novel groups. Averaged over the four known targets, the correlation between the Bradley--Terry ranking and the true ranking is $+0.67$, against $+0.12$ for the three novel ones, a difference of $0.55 \pm 0.27$ ($p = 0.09$, two-sided exact permutation test over the seven targets). Averaging all duels, independent of target, gives a pairwise accuracy of 0.78 on known targets and 0.62 on novel ones ($\dKU = +0.16 \pm 0.07$, $p = 0.11$). The separation is nearly complete and the effect size large, with 11 of the 12 known--novel pairings favoring the known target on both metrics (Cliff's $\delta = 0.83$). With removal of the chemical identity above, the known--novel split does not close either, but sharpens: accuracy becomes 0.75 on known targets and 0.53 on novel ones, $\dKU = +0.23$ ($p = 0.06$, Cliff's $\delta = 1.00$). Dropping temperature is statistically indistinguishable between known and novel materials ($p = 0.31$ and $p = 0.23$). Overall, this suggests being thermally driven being a much more meaningful difference than the known--unknown split.

\textbf{Consistency of the pairwise response.} The rate at which the two presentation orders disagree provides a rough proxy for the model's uncertainty on a target. Every novel material shows a higher flip rate than every known material (0.11 against 0.05, a perfect separation, $p = 0.06$, two-sided exact permutation test), and the flip rate is strongly tied to accuracy across targets ($\rho = -0.79$, $p = 0.048$, exact permutation over all pairings), so the model is least consistent where it is least accurate.

\begin{figure}[t]
  \centering
  \includegraphics[width=1.0\linewidth]{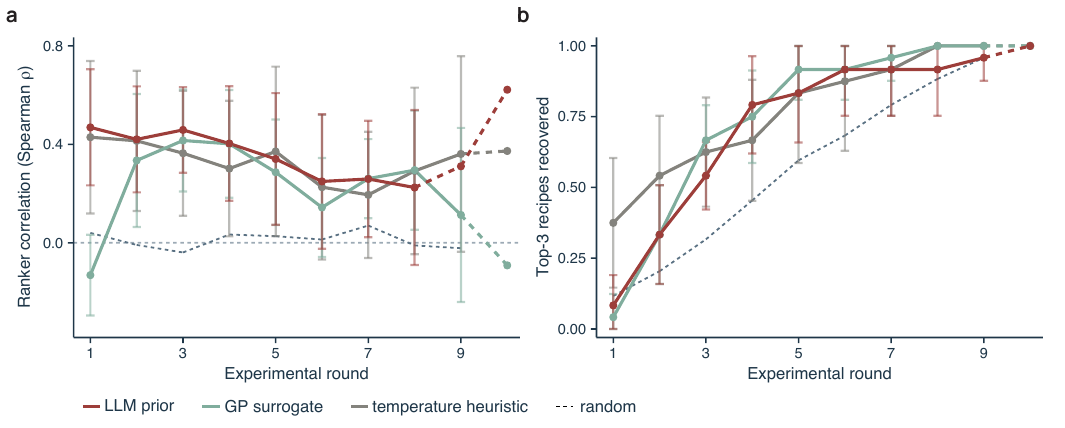}% use vector if possible
    \caption{\textbf{The same rankers replayed as a discovery campaign.} Three recipes are run each round and their measured purities revealed, until the bank is depleted. \textbf{a.} Spearman correlation between a ranker's ordering of the recipes still available and their true purities, per round; round 1 is the cold start, before any measurement exists. \textbf{b.} Fraction of a material's true top three recovered by a given round. Both panels average over the 8 materials with 95\% confidence intervals over materials, and are dashed where fewer than 8 materials still have recipes left.} %The temperature heuristic is a control rather than a deployable strategy, since it presupposes that the target is thermally driven.}
  \label{fig:active}
\end{figure}

\section{Active Learning Results}
\label{sec:al}

\textbf{Ranking as a decision problem.} Since a biased prior can still be a useful one, we assess whether the bias is a liability in a realistic material synthesis campaign, typically judged by how quickly a working recipe is found.

\textbf{Campaign setup.} Following the protocol of Section~\ref{sec:al-methods}, performance is tracked as the per-round Spearman correlation between a ranker's ordering of the remaining recipes and their true purities (Fig.~\ref{fig:active}a, per material in Fig.~\ref{fig:ranker-spearman-round}), as the fraction of the true top three recovered by a given round (Fig.~\ref{fig:active}b, per material in Fig.~\ref{fig:SI-recall-per-material}), and as the ranking efficiency over the whole campaign (Table~\ref{tab:ranker-metrics}). Per-material results are given in Table~\ref{tab:targets}, and the same protocol run with a smaller model in Fig.~\ref{fig:SI-fig3-luna}.

\textbf{The LLM prior against a fitted surrogate.} We compared the two rankers round by round and found that they fail in opposite regimes: the GP has nothing to fit in the opening round and ranks worse than random there ($\rho = -0.13$ against random's $+0.04$), whereas the LLM ranks at $\rho = +0.47$ before any measurement exists. The GP's disadvantage is short-lived: it matches the LLM by round two and overtakes it on recall by round three. Both have recovered 80\% of the true top three by round five. The LLM's advantage lies instead in the quality of the ordering it produces, with a ranking efficiency of 0.74 over the whole campaign against the GP's 0.66 ($p = 0.055$, two-sided exact sign-flip permutation test over the 8 materials).

\textbf{Comparison with an explicit thermal heuristic.} To ask whether a general prior can substitute for knowing the target in advance, we ran the same campaign with the thermal heuristic in place of the model. The LLM proves useful during early discovery but does not outperform that prior knowledge: encoding the expectation that a target is thermally driven yields the strongest early recall of any strategy we ran. Averaged over the whole campaign, the heuristic does not beat the LLM (ranking efficiency 0.72 against 0.74, $p = 0.82$). It also demonstrates the cost of a fixed rule when that rule is wrong. On M1, whose purity falls with temperature, the heuristic reaches a ranking efficiency of 0.22 against random's 0.52 (Table~\ref{tab:SI-ranker-bounds-per-material}), whereas the LLM (0.67) and the GP (0.61) both remain above random. 

\textbf{Ranking with experimental feedback.} To test whether feeding laboratory results back to the LLM recovers what it lacks, we repeated the campaign with outcomes supplied in the prompt. See details and full results in \ref{app:experimental_feedback}. This improved how well the model ordered the untested recipes, raising ranking efficiency from 0.74 to 0.78 ($+0.036 \pm 0.009$, $p = 0.023$, two-sided exact sign-flip permutation test, improving on 7 of 8 materials). However, while showing the model its own results makes it a better ranker, it does not correct its underlying thermal bias.

\textbf{What the campaign establishes.} An LLM prior is useful for roughly the first two campaign rounds, after which a cheap surrogate trained on a few measurements catches up. Establishing this requires physical experiments as we carry out here: published-recipe benchmarks cannot distinguish reasoning from recall, nor evaluate failed syntheses.

%\textbf{What the campaign establishes.} Taken together, an LLM prior is worth approximately the first two rounds of a campaign, after which an inexpensive surrogate fitted to a handful of measurements has caught up. These effects could be established only by carrying out the experiments, since every claim above rests on unsuccessful syntheses and on measured purities for targets absent from any training corpus. Benchmarks assembled from published recipes cannot distinguish a model that reasons about a material from one that has read about it, and they cannot score a recipe that did not work.

\begin{table}[t]
\centering
\small
\caption{\textbf{Ranking metrics for the rankers tested in the active learning loop.} BEDROC \citep{BEDROC_2007} rewards concentrating successful recipes at the very top of a sorted list, and so rewards emptying a bank as well as ordering it; ranking efficiency measures ordering quality alone. The two disagree about the thermal heuristic, which front-loads a few strong recipes but orders the rest poorly. Oracle, anti-oracle and random are reference points. (Ranking efficiency does not reach exactly 1 for the oracle, 0 for the anti-oracle or 0.5 for random, because several recipes share a target phase purity of 0, so no ordering can separate them. For the purposes of mirroring an actual experimental set-up, the recipes with a tied outcome were not removed for the calculation of these metrics.)}
\begin{tabular}{lrr}
\toprule
Ranker & BEDROC & Ranking efficiency \\
\midrule
Oracle: true phase purity ordering & 1.000 & 0.986 \\
LLM prior alone (\texttt{GPT-5.6-Sol}) & 0.379 & \textbf{0.741} \\
GP alone & 0.217 & 0.656 \\
Thermal heuristic alone & \textbf{0.570} & 0.715 \\
Random ranking & 0.261 & 0.506 \\
Anti-oracle: inverted phase purity ordering & 0.000 & 0.014 \\
\bottomrule
\end{tabular}

\label{tab:ranker-metrics}
\end{table}
%\footnotetext{Ranking efficiency does not reach exactly 1 for the oracle, 0 for the anti-oracle or 0.5 for random, because several recipes share a target phase purity of 0, so no ordering can separate them.}

\section{Conclusions}

In this work, we presented an experimental benchmark for solid-state synthesis and used it to evaluate a large language model as a pairwise ranker of recipes. The benchmark itself is the main contribution: 240 measured syntheses covering the same 30 firing conditions on each of eight oxide targets, with purity quantified by X-ray diffraction. Notably, it includes failed attempts and three targets that have never been reported, and are thus guaranteed to be excluded from model pretraining. To our knowledge, it is the first synthesis benchmark built from measured outcomes on a shared condition grid rather than from published recipes.

Alongside the data, a key methodological contribution is the pairwise evaluation design, in which the model only ever judges two recipes of the same target at a time. This allows us to uncover underlying biases in the LLM's chemical decision-making process through information masking. Our central finding is that the chemistry of the target material is not the main driver: the model consistently ranks compounds with thermally-dominated purities higher than those without, and removing the chemistry information has virtually no effect on these targets (pairwise accuracy of 0.74 with chemistry information compared 0.72 without). Chemistry acts as a modulator, occasionally improving ranking performance.

Finally, the presence of strong biases do not render LLMs useless for synthesis planning. By simulating an active learning campaign, we show that an LLM can serve as a 'cold start' prior in cases where no data is available. However, it is nevertheless a prior with strong thermal preference, which may be confidently wrong on materials which are not thermally driven. 

While our benchmark is experimentally grounded, it remains small and its targets are thermally dominated. The extent to which these findings translate to systems governed by, for example, precursor reactivity, will have to be established case by case. Our known--novel comparison is similarly bounded: every metric we measured favors the known group, but eight targets cannot establish the effect with confidence. Resolving it requires further synthesis rather than further analysis.

Ultimately, none of these findings would have been visible in a benchmark built from published recipes, which often contain no failures, no quantitative purities, and likely has been leaked into an LLM's training corpus. In fact, we suspect that many of the underlying biases cannot be uncovered from published materials synthesis recipes. We therefore encourage the community to build experimentally measured synthesis benchmarks and to keep them withheld, since a test set can separate reasoning from recall only for as long as it stays out of the training corpus. The pairwise methodology introduced can be used to probe such benchmarks. We envision that future benchmarks will make the biases of synthesis models measurable rather than assumed, turning the question of whether a model understands chemistry into one that can be settled experimentally.
%We have also examined only prompting; what this bias implies for models fine-tuned on such data remains open, and this benchmark is well placed to address it.

% \section*{References}

\small
\bibliographystyle{plainnat} % or plainnat per your preference/guidelines
\bibliography{refs}

\normalsize

%%%%%%%%%%%%%%%%%%%%%%%%%%%%%%%%%%%%%%%%%%%%%%%%%%%%%%%%%%%%

\newpage
\appendix

\section{Technical Appendices and Supplementary Material}

\subsection{Related work on injecting priors into synthesis prediction}
\label{app:priors}

Several lines of work try to improve synthesis prediction by supplying the model with information it would not otherwise have. Thermodynamic estimation narrows the set of enumerated reactions by selectivity and tracks which impurity phases are expected to appear \citep{McDermott_2023}, and can be used to judge predictions where experimental ground truth is missing \citep{schlesinger2026}. Recipes for chemically related materials improve generalization and can be retrieved under thermodynamic filtering, ranked in a shared representation, or generated to augment pretraining \citep{noh2025, prein2025retrorankin, Prein2025SyntMTE}. Experimental outcomes can be fed back during search \citep{Szymanski2023_ARROWS3}, and without such a loop that information is simply lost \citep{Prein2025Pipeline}. Others restrict the output format, for example through fixed intermediate representations \citep{zheng2023chatgpt}, following the documentation practice established for models and datasets \citep{mitchell2019model, gebru2021datasheets}.

Each channel is bounded. Thermodynamic control requires the target to exceed all competing phases by at least 60\,meV per atom, which holds for roughly 15\% of enumerated reactions \citep{szymanski2024quantifying}, while literature transfer requires a close analogue and cannot leave its training vocabulary \citep{noh2025}. The mapping from target to synthesis is one-to-many, so a single supplied route misrepresents the problem \citep{pan2026diffsyn}. Supplied information is not neutral either: models override their own correct prior with incorrect retrieved content in more than 60\% of cases \citep{wu2024clasheval}. Whether injected knowledge helps or misleads is therefore an empirical question, and one that remains open, since the three channels have never been compared on the same targets.

\subsection{X-ray diffraction fitting and phase quantification}
\label{app:xrd}

To extract material purity, we used an in-house phase-identification algorithm that picked likely phases from the ICSD, Materials Project, GNoME, and Alexandria. These were weighted based on thermodynamic likelihood, scanned for peak similarity, and finally fitted by an automated Rietveld refinement procedure performed using BGMN \citep{doebelin2015profex}, a robust automated Rietveld refinement algorithm. All results were then human-validated by a diffraction specialist, including against alternative phase fits. The reported purities are wt-\% from Rietveld refinement and carry a phase-dependent 1-2\%. Non-crystalline phases are also fitted; none of the targets in this benchmark form fully amorphous phases in this experimental set-up.

As we cannot reveal the target materials, we also cannot reveal all XRD fits. An example fit of Material 7 is given below in Fig. \ref{fig:xrd} to showcase a typical XRD fit. The phase fits are shown in Table \ref{tab:xrd}.

\begin{table}[ht]
\centering
\caption{Li$_7$La$_3$Zr$_2$O$_{12}$, $R_{\mathrm{wp}} = 4.87$, GoF = 1.14)}
\begin{tabular}{lr}
\hline
\textbf{Phase} & \textbf{wt\%} \\
\hline
Li$_7$La$_3$Zr$_2$O$_{12}$ \textit{(target)} & \textbf{91.4} \\
La$_2$Zr$_2$O$_7$ & 5.5 \\
Li$_2$ZrO$_3$ & 3.1 \\
\hline
\label{tab:xrd}
\end{tabular}
\end{table}

\begin{figure}[ht]
  \centering
  \includegraphics[width=0.7\linewidth]{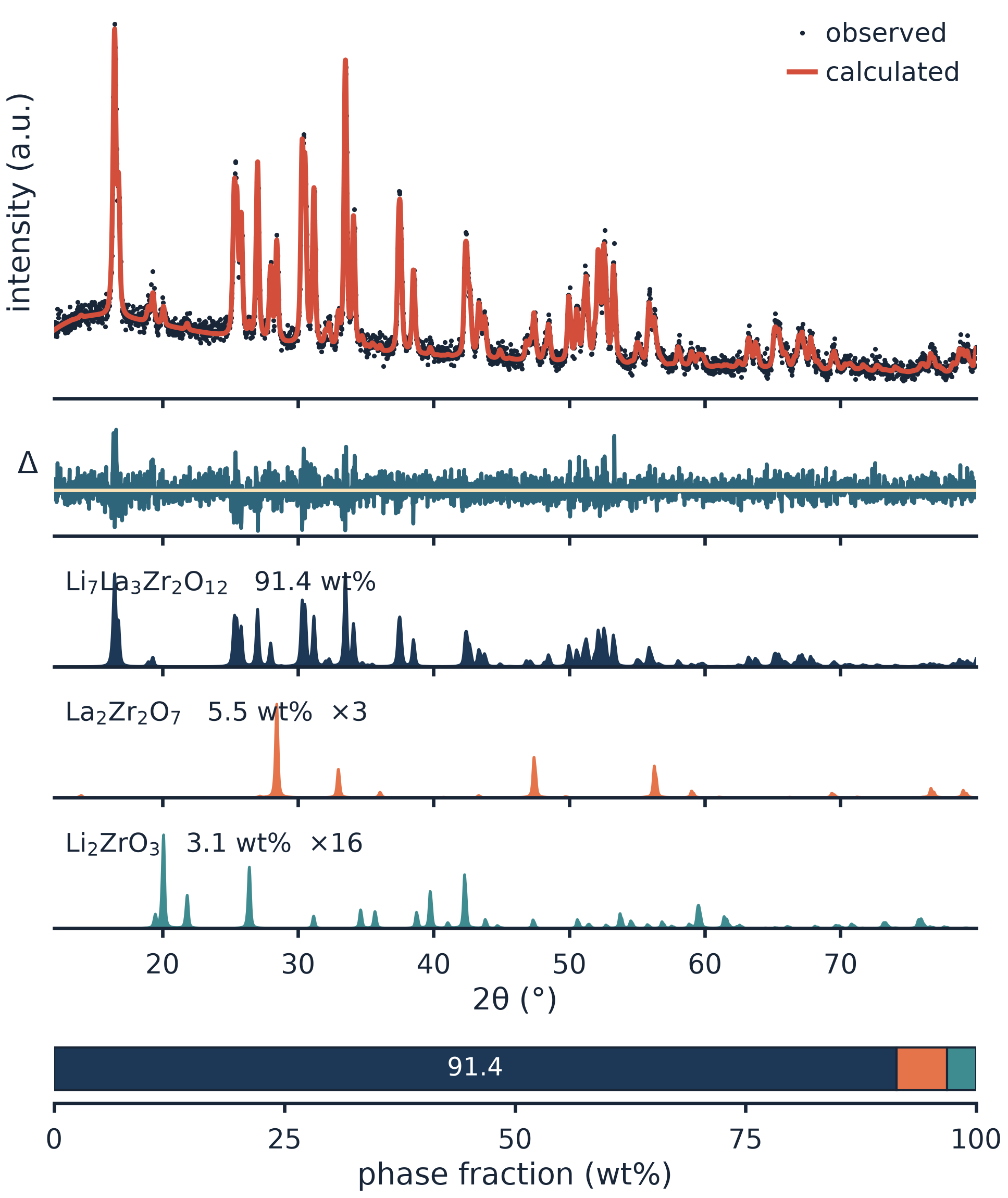}% use vector if possible
    \caption{\textbf{XRD fit example.} Li$_7$La$_3$Zr$_2$O$_{12}$ fit, with weight-\% (wt\%) by fitted phase. La$_2$Zr$_2$O$_7$'s peaks are shown with 3x height for comparison and Li$_2$ZrO$_3$ with 16x. $R_{\mathrm{wp}} = 4.87$, GoF = 1.14}
  \label{fig:xrd}
\end{figure}

\subsection{Bradley--Terry fitting}
\label{app:bt}

Strengths are fit by minimizing
$\mathcal{L}(\theta) = \sum_{(w,l) \in \mathcal{D}} -\log \sigma(\theta_w - \theta_l) + \frac{\lambda}{2}\lVert\theta\rVert_2^2$
over all observed pairwise outcomes $\mathcal{D}$ (winner $w$, loser $l$), with $\lambda = 10^{-3}$, via L-BFGS-B. One recipe's strength is pinned to $0$ for identifiability, since the model is invariant to a constant shift in $\theta$, and the fitted scores are then mean-centered.

\subsection{Duel scoring}
\label{app:scoring}

For duel $d = (r, r')$ let $y_d = \mathbb{1}[\,p(r) > p(r')\,]$ indicate that the first-presented recipe is the purer one, and $\hat y_d = \mathbb{1}[\,J(\phi(r), \phi(r')) = A\,]$ that the judge named the first. The duel scores $s_d = \mathbb{1}[\,\hat y_d = y_d\,] \in \{0, 1\}$, defined only when $J \in \{A, B\}$. Per-target accuracy is then
$$A_c(t) = \frac{1}{|D_t^{\ast}|} \sum_{d \in D_t^{\ast}} s_d ,
\qquad D_t^{\ast} = \{ d \in D_t : \delta_d > 0,\ s_d \text{ defined} \} .$$

\textbf{Reporting conventions.} Group means pool over recipe pairs, whereas every significance test permutes the seven target labels, since pairs within a target share its chemistry, its bank and its ceiling. Reported intervals are standard errors: a bootstrap over targets for group differences, and the standard error of the mean over per-target paired differences for ablation changes.

\subsection{BEDROC}
\label{app:bedroc}

With $r_{\text{a}} = n/N$, $N$ the size of the material's bank, $n$ the number of active candidates and $r_i$ the rank of the $i$-th success,
$$\mathrm{BEDROC} = \mathrm{RIE} \cdot \frac{r_{\text{a}} \sinh(\alpha/2)}{\cosh(\alpha/2) - \cosh(\alpha/2 - \alpha r_{\text{a}})} + \frac{1}{1 - e^{\alpha(1 - r_{\text{a}})}}, \qquad
\mathrm{RIE} = \frac{\sum_{i=1}^{n} e^{-\alpha r_i / N}}{\frac{n}{N}\cdot\frac{1 - e^{-\alpha}}{e^{\alpha/N} - 1}}.$$

\subsection{Duel identifiers}
\label{app:duelid}

A duel is identified by $\mathrm{id} = H(t, r_{\text{lo}}, r_{\text{hi}}, r_{\text{first}})$ with $H$ a fixed hash and $r_{\text{lo}}, r_{\text{hi}}$ the lower- and higher-purity member of the pair. The identifier does not depend on the prompt condition, so every condition is evaluated on exactly the same duel set and any two conditions can be differenced pair by pair.

\subsection{Gaussian process surrogate}
\label{app:gp}

In the first round, when no data has been acquired, a deterministic algorithm prioritizing variety selects the first 3 recipes: recipes are greedily selected to satisfy, in order, a minimum count fired at the temperature extremes of the accessible range (lowest and highest) and a minimum count of refire routes, before the remaining slots of the round are filled in a fixed order. This ensures the earliest measurements span the accessible processing conditions rather than clustering on a single regime, before there is any fitted model to guide selection.

At the start of each subsequent round the GP is fitted on 5 features of the measured recipes: peak firing temperature, total dwell time, refiring conditions (binary, whether the temperature profile contains more than one firing step), atmosphere conditions (binary, whether an air atmosphere was used at any point in the profile, as opposed to vacuum), and thermal budget, $\mathrm{TB} = \sum_{s} T_s t_s$, summed over every segment $s$ of the temperature profile, including heating ramps, dwells and cooling, where $T_s$ is the segment's target temperature and $t_s$ its duration. We use the settings proposed by \citet{chen_2026} for a GP based on a Mat\'ern-5/2 kernel with automatic relevance determination (ARD) and a dimension-aware hyperprior placed on both the lengthscale, $\ell$, and outputscale parameters.

\subsection{Known-unknown accuracy split}
\label{app:per_target}

A target cannot produce a gap wider than its own best recipe, so the groups are comparable only up to $|\dpT| \approx 0.35$, beyond which the novel group loses M2, then M1, and is eventually represented by M3 alone.

\begin{figure}[ht]
  \centering
  \includegraphics[width=1.0\linewidth]{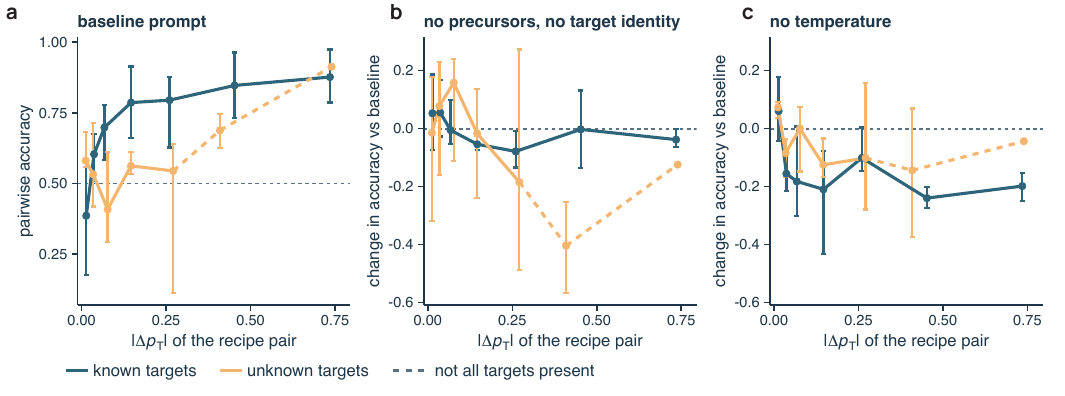}% use vector if possible
    \caption{\textbf{Pairwise recipe ranking performance for known and novel materials.} Accuracy resolved by the purity gap $|\dpT|$, for known (n=4) and novel (n=3) targets. A target cannot produce a gap wider than its own best recipe, so the widest bins lose targets from the novel group; those stretches are dashed, and the solid stretch is where the groups are comparable. \textbf{a.} Full prompt. \textbf{b.} Target identity and precursors removed. \textbf{c.} All temperature information removed. Panels b and c show the change against panel a. Per target curves are shown in Fig. \ref{fig:per_target_pairwise}.}
  \label{fig:ku_discrimination}
\end{figure}

\subsection{Ranking with experimental feedback}
\label{app:experimental_feedback}

\textbf{Information given.} Alongside the pair being judged, it carried the top 3 and bottom 2 recipes discovered so far, described by peak firing temperature, total furnace time, atmosphere, and the presence of a refiring step, together with their measured purities. Table~\ref{tab:ranker-metrics-feedback} and Fig.~\ref{fig:SI-fig3-LLM-feedback} show additional information.

This information improved how well the model ordered the untested recipes, raising ranking efficiency from 0.74 to 0.78 in the active-learning set-up (+0.036 ± 0.009, p = 0.023, two-sided exact sign-flip permutation test, improving on 7 of 8 materials). The one material whose ranking showed no effect after the LLM was provided with  experimental feedback was M1, the one target for which the bias points in the wrong direction, which barely moved ($+0.004$ against a mean of $+0.036$).

% generated by scripts/experiments/datacard-fapo/report_ranker_comparison_topk3_nocard.py (sol)
\begin{table}[H]
\centering
\small
\begin{tabular}{lrr}
\toprule
Ranker & BEDROC & Ranking efficiency \\
\midrule
Oracle: true $\pT$ ordering & 1.000 & 0.986 \\
LLM prior alone (\texttt{GPT-5.6-Sol}) & 0.379 & 0.741 \\
LLM $+$ outcome feedback (top-3/worst-2 each round) & 0.457 & \textbf{0.776} \\
GP alone & 0.217 & 0.656 \\
Thermal heuristic alone & \textbf{0.570} & 0.715 \\
Random ranking & 0.261 & 0.506 \\
Anti-oracle: inverted $\pT$ ordering & 0.000 & 0.014 \\
\bottomrule
\end{tabular}

\caption{\textbf{Ranking metrics with outcome feedback added.} The same campaign as Table~\ref{tab:ranker-metrics}, with an additional arm in which the model is shown the top 3 and bottom 2 recipes discovered so far, alongside their measured purities, in the prompt for every duel.}
\label{tab:ranker-metrics-feedback}
\end{table}

\subsection{Additional controls by information removal}
\label{app:controls}

We provide the results for different information removal tested on \texttt{GPT-5.6-Sol} and \texttt{GPT-5.6-Luna} below. Mask refers to information that is withheld from the model. The base prompt used has minimal information and is given in \ref{app:prompt}. A minimal prompt was found to be optimal as shown in \ref{app:prompt_opt}. Note that in the case that the target is masked, the prompt still has access to the precursors, but not the relative amounts. In the case the precursors are hidden, the target composition is still visible. Removing both removes all references to the target (in effect converting the measure to a pure "which profile is better?" measure).

\begin{table}[ht]
\centering
\scriptsize
\resizebox{\textwidth}{!}{%
\begin{tabular}{lcccccccc}
\toprule
 & M1 & M2 & M3 & M4 & M5 & M6 & M7 & M8 \\
\midrule
group
& unknown & unknown & unknown & excluded & known & known & known & known \\

\texttt{GPT-5.6-Sol}
& 0.62 & 0.45 & 0.71 & 0.73 & \textbf{0.75} & 0.86 & 0.66 & 0.85 \\

\texttt{GPT-5.6-Sol}\_no\_target/precursors 
& 0.25 & 0.57 & 0.68 & 0.76 & 0.71 & 0.81 & 0.69 & 0.81 \\

\texttt{GPT-5.6-Sol}\_no\_temperature
& 0.38 & 0.43 & 0.71& 0.44 & 0.63 & 0.71 & 0.50 & 0.57 \\

\texttt{GPT-5.6-Sol}\_no\_atmosphere
& \textbf{0.67} & 0.51 & 0.42 & 0.76 & 0.70 & 0.81 & 0.68 & 0.83 \\

\texttt{GPT-5.6-Luna}
& 0.48 & 0.52 & \textbf{0.73} & \textbf{0.78} & \textbf{0.75} & \textbf{0.87} & 0.68 & \textbf{0.88} \\

\texttt{GPT-5.6-Luna}\_no\_target/precursors 
& 0.25 & 0.57 & 0.69 & 0.71 & 0.70 & 0.78& 0.66 & 0.77 \\

\texttt{GPT-5.6-Luna}\_no\_temperature
& 0.38 & 0.46 & 0.69 & 0.54 & 0.61 & 0.72 & 0.50 & 0.58 \\

\texttt{GPT-5.6-Luna}\_no\_atmosphere
& 0.43 & \textbf{0.62} & 0.53 & 0.76 & 0.67 & 0.82 & \textbf{0.71} & 0.85 \\

\bottomrule
\end{tabular}%
}
\caption{Per-target accuracy across different conditions and models. No target/precursors means that the chemistry is hidden to the model, while no temperature hides the max temperature and the thermal budget. Note that \texttt{Sol} is used throughout due to its highest performance on unknown materials, see below in Table \ref{tab:masking_summary}}
\label{tab:per_target_accuracy}
\end{table}

\begin{table}[ht]
\centering
\scriptsize
\resizebox{\textwidth}{!}{%
\begin{tabular}{llcccccc}
\toprule
Model & Mask & Acc. total & Acc. known & Acc. unknown &
$\Delta_{KU}$ & Picks hotter & Flip rate \\
\midrule

\texttt{Sol} & --
& 0.72 & 0.78 & \textbf{0.62} & 0.16 & 0.72 & 0.09 \\

\texttt{Sol} & Precursors
& 0.72 & 0.78 & \textbf{0.62} & 0.16 & 0.71 & 0.09 \\

\texttt{Sol} & Target
& 0.66 & 0.68 & 0.56 & 0.12 & 0.68 & 0.12 \\

\texttt{Sol} & Target + precursors
& 0.68 & 0.75 & 0.53 & 0.23 & 0.88 & \textbf{0.06} \\

\texttt{Sol} & Max. temp. + thermal budget
& 0.56 & 0.60 & 0.54 & 0.06 & 0.55 & 0.11 \\

\texttt{Sol} & Atmosphere
& 0.68 & 0.75 & 0.52 & 0.24 & 0.70 & 0.11 \\

\texttt{Sol} & All process fields
& 0.49 & 0.44 & 0.49 & -0.05 & 0.48 & 0.46 \\

\midrule

\texttt{Luna} & --
& \textbf{0.73} & \textbf{0.80} & 0.60 & 0.20 & 0.78 & 0.11 \\

\texttt{Luna} & Precursors
& 0.72 & 0.78 & 0.60 & 0.19 & 0.82 & 0.12 \\

\texttt{Luna} & Target
& 0.71 & 0.78 & 0.56 & 0.21 & 0.82 & 0.13 \\

\texttt{Luna} & Target + precursors
& 0.66 & 0.73 & 0.53 & 0.20 & 0.79 & 0.15 \\

\texttt{Luna} & Thermal budget
& 0.72 & 0.79 & 0.58 & 0.21 & 0.79 & 0.13 \\

\texttt{Luna} & Max. temp.
& 0.64 & 0.68 & 0.57 & 0.10 & 0.67 & 0.11 \\

\texttt{Luna} & Max. temp. + thermal budget
& 0.58 & 0.60 & 0.54 & 0.06 & 0.55 & 0.14 \\

\texttt{Luna} & Dwell time
& 0.72 & 0.79 & 0.59 & 0.20 & 0.82 & 0.12 \\

\texttt{Luna} & Atmosphere
& 0.69 & 0.76 & 0.53 & 0.23 & 0.80 & 0.12 \\

\bottomrule
\end{tabular}%
}
\caption{Accuracy and behavioural effects under feature masking.
$\Delta_{KU} = \mathrm{Acc}_{\mathrm{known}} - \mathrm{Acc}_{\mathrm{unknown}}$. 'Picks hotter' is the rate at which the model picks the hotter recipe, and flip rate the rate at which A then B is different from B then A in the pair-pair comparison. 'All process fields' is statistically equivalent to a coin flip.}
\label{tab:masking_summary}
\end{table}

\subsection{\texttt{GPT-5.6-Sol} Prompt}
\label{app:prompt}
\begin{ttfamily}
\textbf{System}
You are a scientific assistant specialized in solid-state synthesis. You compare two synthesis recipes for the same target material and choose which one is more likely to produce higher target phase purity. You must return exactly one JSON object and nothing else.

User
Target material: Li7La3Zr2O12

The two recipes are shown in random order.
Do NOT assume recipe A is better because it appears first.

Recipe A
Target: Li7La3Zr2O12

Top-level process features:\\
- max\_temp\_c: 1100.0\\
- dwell\_at\_max\_temp\_h: 8.000\\
- thermal\_budget: 17666.667

Precursors:\\
- Li2CO3\\
- La2O3\\
- ZrO2

Process summary:\\
- n\_milling\_steps: 2\\
- milling\_energy\_proxy: 1530000.0\\
- n\_furnace\_steps: 1

Furnace profile:\\
- furnace step 1:\\
  - atmosphere: air\\
  - max\_temp\_c: 1100.0\\
  - dwell\_at\_max\_temp\_h: 8.000\\
  - total\_ramp\_count: 5

Recipe B
Target: Li7La3Zr2O12

Top-level process features:\\
- max\_temp\_c: 550.0\\
- dwell\_at\_max\_temp\_h: 4.000\\
- thermal\_budget: 3283.333

Precursors:\\
- Li2CO3\\
- La2O3\\
- ZrO2

Process summary:\\
- n\_milling\_steps: 2\\
- milling\_energy\_proxy: 1530000.0\\
- n\_furnace\_steps: 1

Furnace profile:\\
- furnace step 1:\\
  - atmosphere: vacuum\\
  - max\_temp\_c: 550.0\\
  - dwell\_at\_max\_temp\_h: 4.000\\
  - total\_ramp\_count: 5

Choose the recipe more likely to produce higher target phase purity.

Output rules:
- Return exactly one line of valid JSON.
- Return no explanation, no analysis, no markdown, and no extra text.
- Valid outputs are only:
{"winner":"A"}
{"winner":"B"}
- Any other output is invalid.
\end{ttfamily}

\subsection{Prompt Optimisation}
\label{app:prompt_opt}

We also optimized the information given to the model. Here, we used \texttt{GPT-5.6-Luna} due to the lower cost. The minimal prompt case is what we used in later benchmarking. Guided scaffold included explicit guidance on considering the following:

\begin{ttfamily}
 
Evaluation principles:\\
- Rank recipes by relative likelihood of obtaining predominantly single-phase Li7La3Zr2O12.\\
- Prioritize temperature-driven kinetics and thermal budget over minor nominal stoichiometric deviations.\\
- Do not infer specific secondary phases solely from small stoichiometric offsets.\\
- Treat identical or poorly specified parameters as weakly discriminative.

Compare the two recipes using this checklist as guidance:\\
- Stoichiometric plausibility\\
- Temperature profile\\
- Dwell time at high temperature\\
- Milling / homogenization\\
- Cooling conditions\\
- Ramp rates

Focus only on the most discriminative factors for this pair.
   
\end{ttfamily} 

\vspace{1cm}

with variations adding only part of this information. See below in Table \ref{tab:luna_prompt_optimization}. 

We also changed the prompt, suggesting various additional pieces of information or hints for the model. On average, none of these improved performance, with most degrading performance or having no effect, while increasing the rate of the model picking the hotter recipe. The minimal prompt was the best performing on all metrics except accuracy on unknown targets, where it was 0.01 worse (0.60 vs 0.61) than two other prompts (Table \ref{tab:luna_prompt_optimization}). As a result, we continued with a relatively "bare" prompt in further benchmarking. 

The minimal prompt was chosen using accuracy measured on this benchmark rather than on held-out data; we performed that selection on \texttt{GPT-5.6-Luna} and transferred the resulting prompt unchanged to \texttt{GPT-5.6-Sol}; note that overall accuracy varies by only 0.02 across all fourteen variants (0.71–0.73), whereas the rate at which the model picks the hotter recipe varies from 0.65 to 0.95.

\begin{table}[ht]
\centering
\scriptsize
\resizebox{\textwidth}{!}{%
\begin{tabular}{llcccccc}
\toprule
Type & Prompt modification & Acc. total & Acc. known & Acc. unknown &
$\Delta_{KU}$ & Picks hotter & Flip rate \\
\midrule

Baseline & Minimal prompt
& \textbf{0.73} & \textbf{0.80} & 0.60
& 0.20 & 0.78 & 0.11 \\

\midrule

Guidance & Guided scaffold
& 0.71 & 0.78 & 0.54
& 0.24 & 0.95 & 0.06 \\

Guidance & Minus thermal-priority line
& 0.72 & 0.79 & 0.57
& 0.23 & 0.80 & 0.13 \\

Guidance & Minus thermal line + budget field
& 0.71 & 0.79 & 0.56
& 0.23 & 0.82 & 0.12 \\

Guidance & Minus stoichiometry rule
& 0.71 & 0.79 & 0.55
& 0.24 & 0.95 & \textbf{0.05} \\

Guidance & Minus both chemistry rules
& \textbf{0.73} & \textbf{0.80} & 0.59
& 0.21 & 0.78 & 0.13 \\

\midrule

Add & Full furnace schedule
& 0.72 & 0.79 & 0.59
& 0.19 & 0.74 & 0.14 \\

Add & Precursor amounts
& 0.72 & 0.79 & 0.57
& 0.23 & 0.81 & 0.12 \\

Add & Long brief + full schedule
& \textbf{0.73} & 0.79 & \textbf{0.61}
& 0.18 & 0.74 & 0.13 \\

Add & JSON on literature information
& 0.72 & 0.79 & 0.59
& 0.20 & 0.73 & 0.11 \\

Hint & Focus on the target
& \textbf{0.73} & \textbf{0.80} & 0.59
& 0.21 & 0.81 & 0.10 \\

Hint & Imagine the ideal recipe
& \textbf{0.73} & 0.79 & 0.60
& 0.19 & 0.73 & 0.12 \\

Hint & Reason over precursor reactions
& \textbf{0.73} & 0.79 & \textbf{0.61}
& 0.18 & 0.72 & 0.10 \\

Hint & Reason on precursor + temperature
& 0.71 & 0.77 & 0.59
& 0.19 & 0.65 & 0.13 \\

\bottomrule
\end{tabular}%
}
\caption{Prompt-optimization results for Luna.
$\Delta_{KU} = \mathrm{Acc}_{known} - \mathrm{Acc}_{unknown}$.
Bold indicates the highest displayed accuracy in each accuracy column and the lowest displayed flip rate.}
\label{tab:luna_prompt_optimization}
\end{table}

\subsection{Heuristic agreement}
\label{app:agreement}

\begin{figure}[H]
  \centering
  \includegraphics[width=1.0\linewidth]{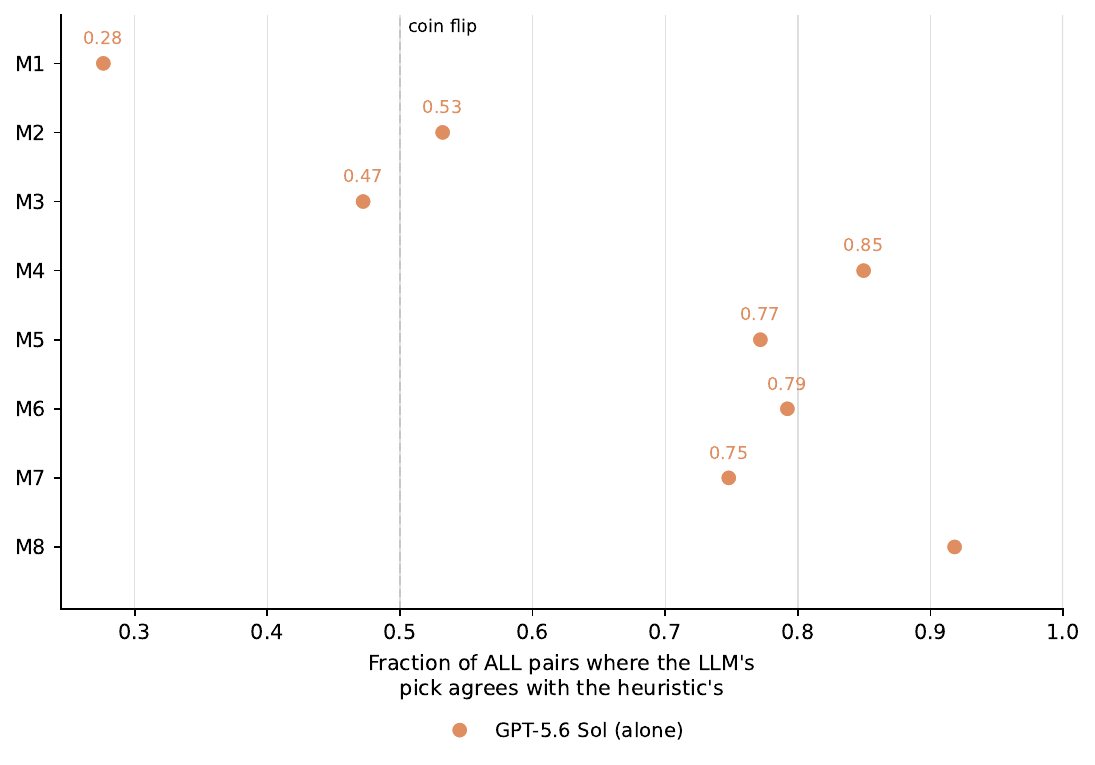}%
    \caption{Fraction of pairwise recipe comparisons where the LLM's pick agrees with the thermal heuristic's, across the 8 benchmark materials. Agreement is 70\%, pooled over all scorable comparisons, and exceeds three quarters on M4, M5, M6 and M8.}
  \label{fig:SI-heuristic-agreement}
\end{figure}

\subsection{Models Probed}

The probed \texttt{GPT-5.6-Sol} version was 2026-06-23 15:20:18 UTC (epoch 1782228018).

The probed \texttt{GPT-5.6-Luna} version was 2026-06-23 15:30:58 UTC (epoch 1782228658).

\subsection{Computational Budget}

The CPU and GPU budget to run the GP and other models was negligible. A total of 1.528B input tokens across 255.7K requests was used to run the LLM benchmarks and develop the prompt.

\newpage

\subsection{Per-material ranking metrics}

% generated by figures/Fig-SI/table_per_material/table_per_material.py
\begin{table}[H]
\centering
\resizebox{\textwidth}{!}{%
\begin{tabular}{lrrrrrrrrrr}
\toprule
 & \multicolumn{2}{c}{LLM prior alone} & \multicolumn{2}{c}{GP alone} & \multicolumn{2}{c}{Temp/dwell heuristic} & \multicolumn{2}{c}{LLM $+$ outcome feedback} & \multicolumn{2}{c}{Random ranker} \\
\cmidrule(lr){2-3} \cmidrule(lr){4-5} \cmidrule(lr){6-7} \cmidrule(lr){8-9} \cmidrule(lr){10-11}
Material & BEDROC & Rank. eff. & BEDROC & Rank. eff. & BEDROC & Rank. eff. & BEDROC & Rank. eff. & BEDROC & Rank. eff. \\
\midrule
M1 & 0.331 & 0.665 & 0.784 & 0.605 & 0.001 & 0.224 & 0.287 & 0.669 & 0.172 & 0.517 \\
M2 & 0.000 & 0.406 & 0.000 & 0.546 & 0.001 & 0.597 & 0.000 & 0.477 & 0.068 & 0.514 \\
M3 & 0.168 & 0.785 & 0.060 & 0.778 & 0.849 & 0.719 & 0.642 & 0.841 & 0.394 & 0.517 \\
M4 & 0.307 & 0.795 & 0.053 & 0.631 & 0.773 & 0.871 & 0.203 & 0.853 & 0.284 & 0.508 \\
M5 & 0.926 & 0.829 & 0.131 & 0.663 & 0.924 & 0.788 & 0.916 & 0.825 & 0.301 & 0.480 \\
M6 & 0.590 & 0.866 & 0.043 & 0.704 & 0.733 & 0.881 & 0.617 & 0.902 & 0.178 & 0.515 \\
M7 & 0.410 & 0.702 & 0.602 & 0.605 & 0.350 & 0.727 & 0.442 & 0.744 & 0.491 & 0.474 \\
M8 & 0.301 & 0.878 & 0.059 & 0.716 & 0.933 & 0.909 & 0.549 & 0.901 & 0.195 & 0.524 \\
\bottomrule
\end{tabular}%
} % Closes the resizebox
\caption{\textbf{Per-material breakdown of BEDROC and ranking efficiency of the different rankers.}}
\label{tab:SI-ranker-bounds-per-material}
\end{table}

\subsection{Outcome feedback and model comparison}
\label{app:models}

\begin{figure}[H]
  \centering
  \includegraphics[width=1.0\linewidth]{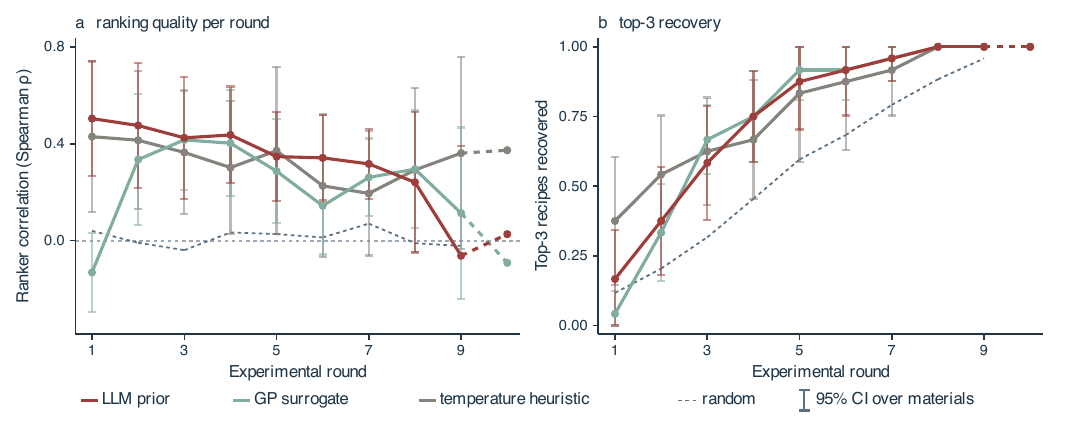}% use vector if possible
    \caption{The campaign of Fig.~\ref{fig:active} run with \texttt{GPT-5.6-Luna}, a smaller model, as the pairwise judge. \textbf{a.} Ranker correlation per round. \textbf{b.} Fraction of the true top three recovered by a given round.}
  \label{fig:SI-fig3-luna}
\end{figure}

\begin{figure}[H]
  \centering
  \includegraphics[width=1.0\linewidth]{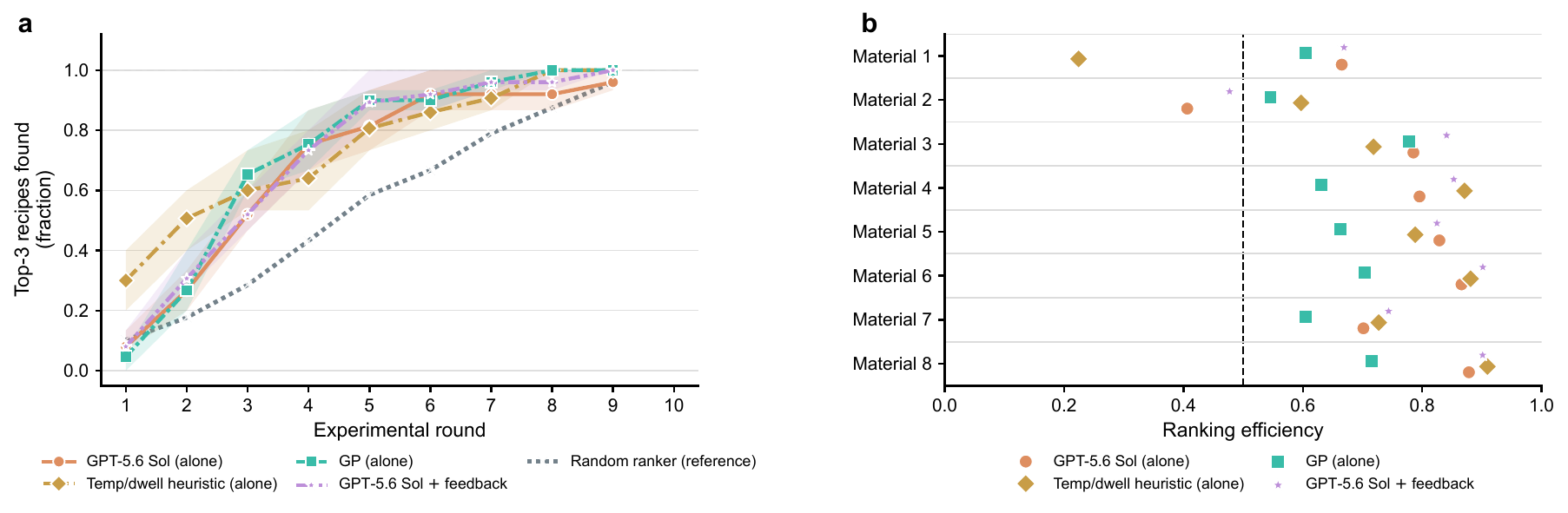}%
    \caption{The campaign of Fig.~\ref{fig:active} with an additional arm in which the LLM is given outcome feedback in the prompt: the top 3 best and bottom 2 worst measured recipes discovered so far, with their target phase purities.}
  \label{fig:SI-fig3-LLM-feedback}
\end{figure}

\begin{figure}[H]
  \centering
  \includegraphics[width=1.0\linewidth]{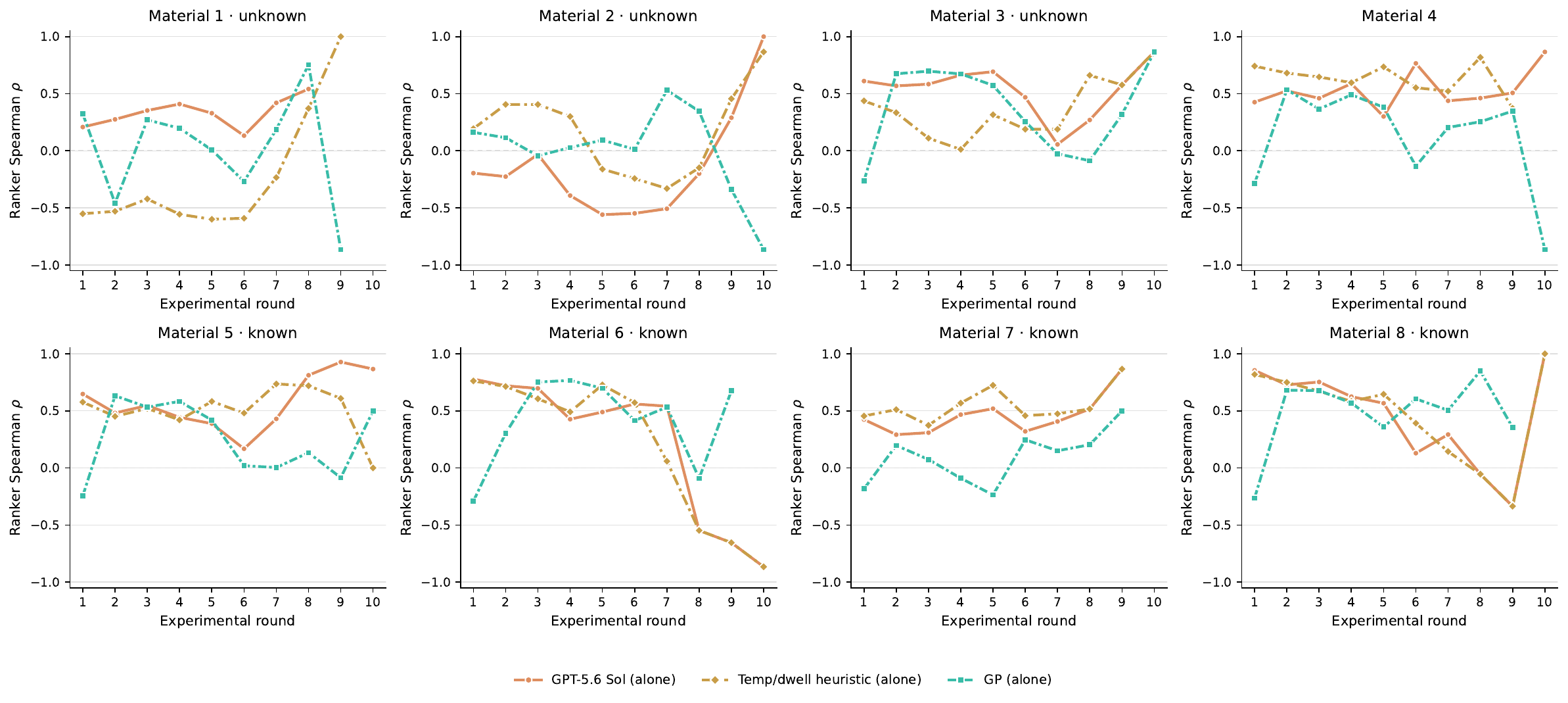}%
    \caption{Per-material ranker Spearman correlation across experimental rounds, as the pool is depleted by running the top 3 ranked recipes each round. LLM = \texttt{GPT-5.6-Sol}.}
  \label{fig:ranker-spearman-round}
\end{figure}

\begin{figure}[H]
  \centering
  \includegraphics[width=1.0\linewidth]{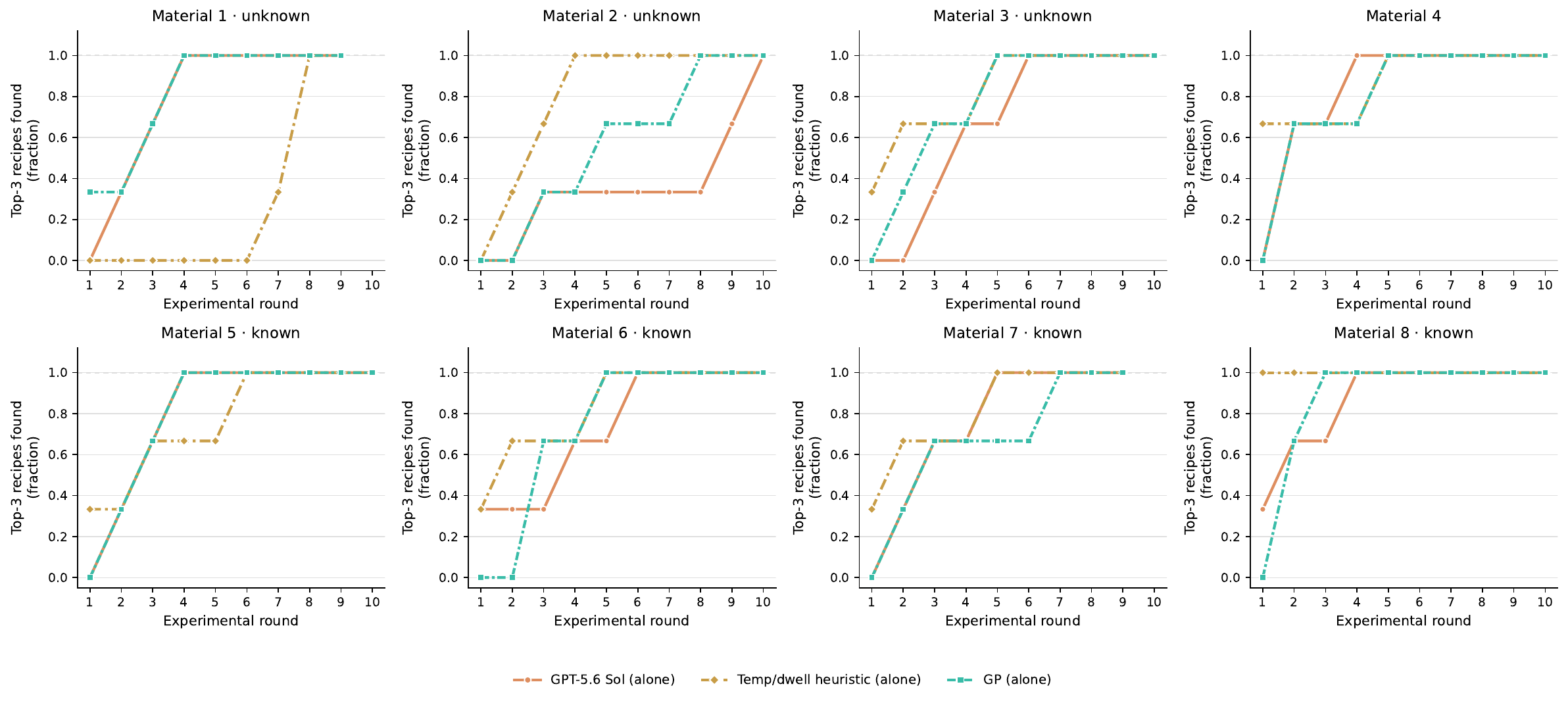}%
    \caption{Per-material fraction of the true top-3 recipes recovered across campaign rounds. LLM = \texttt{GPT-5.6-Sol}.}
  \label{fig:SI-recall-per-material}
\end{figure}

\begin{figure}[H]
  \centering
  \includegraphics[width=1.0\linewidth]{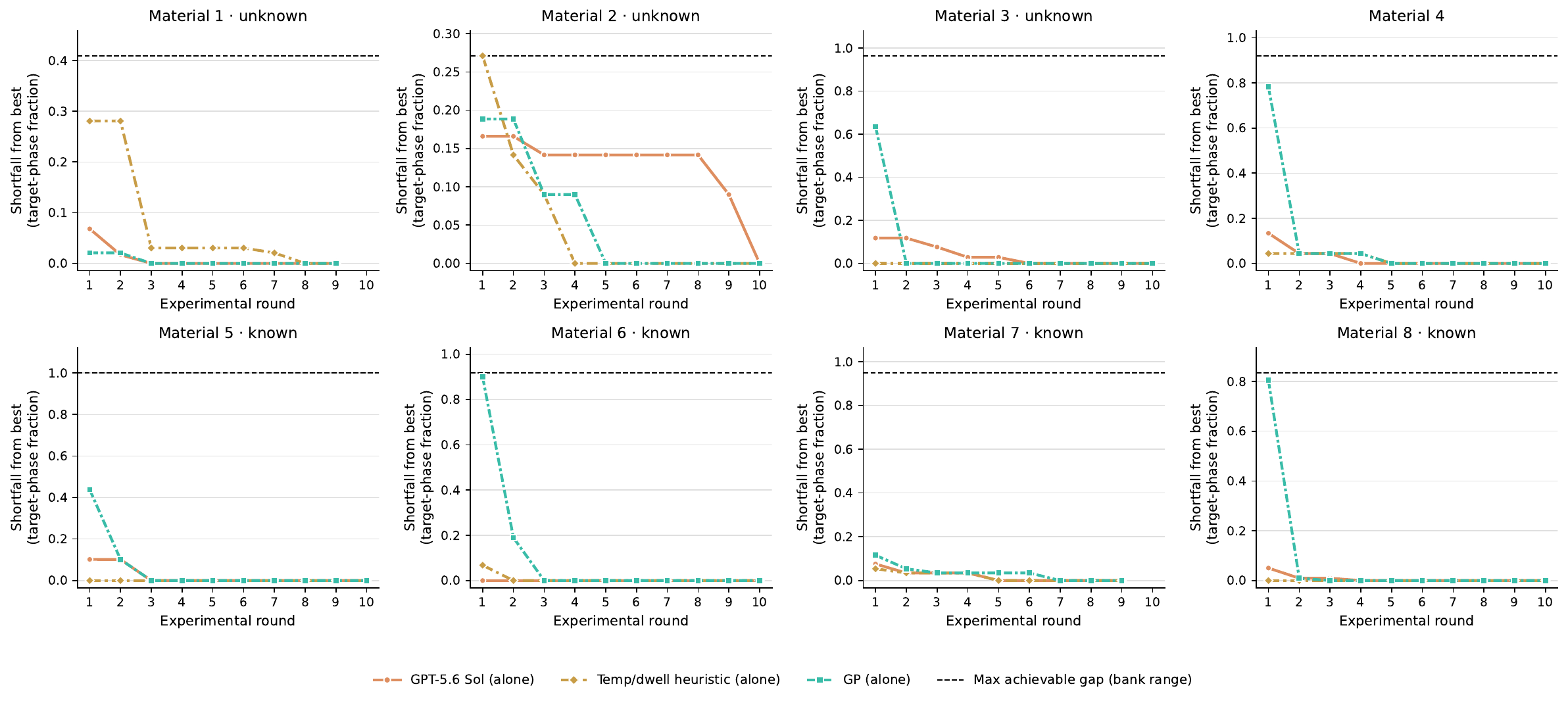}%
    \caption{Per-material simple regret, the shortfall from the best available target phase purity, across campaign rounds. LLM = \texttt{GPT-5.6-Sol}.}
  \label{fig:SI-regret-per-material}
\end{figure}

\begin{figure}[H]
  \centering
  \includegraphics[width=1.0\linewidth]{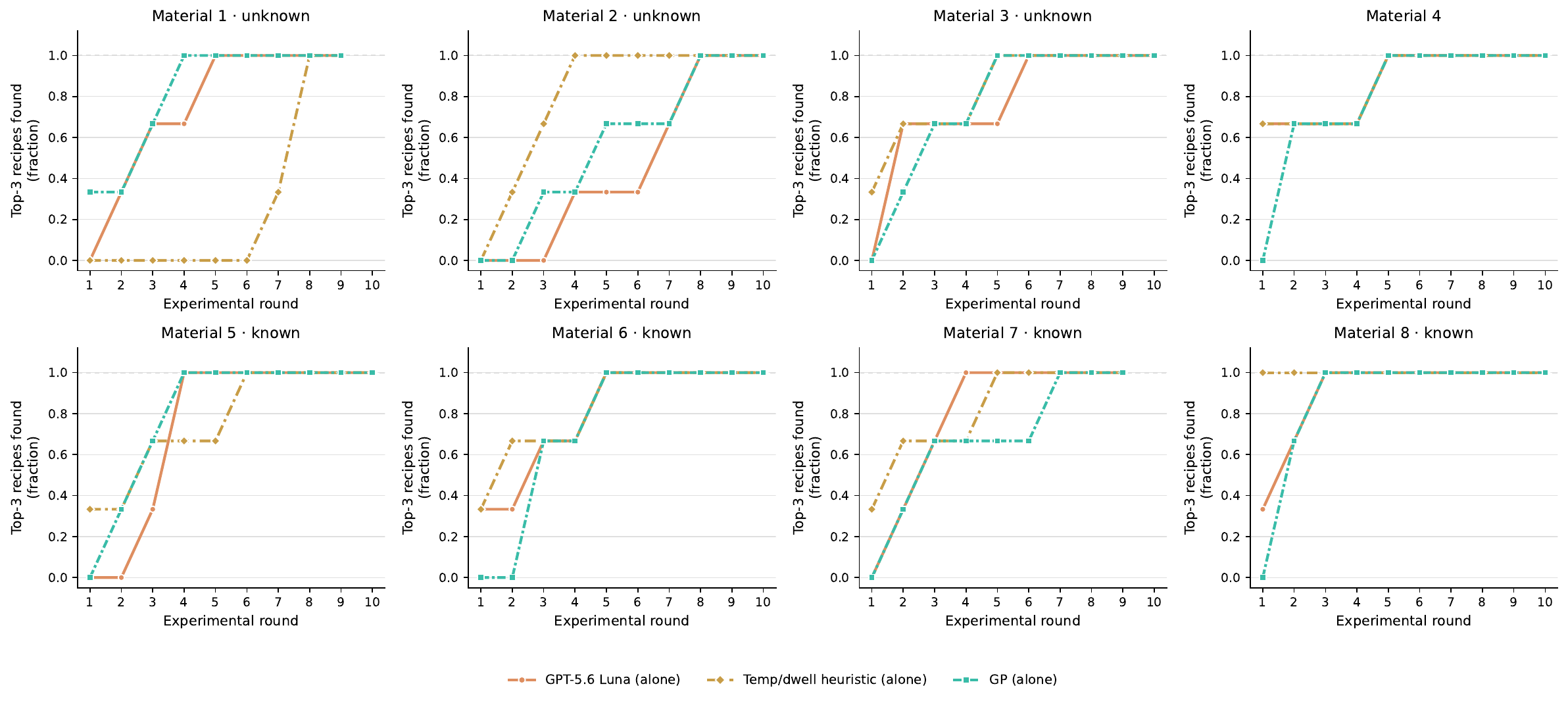}%
    \caption{Per-material fraction of the true top-3 recipes recovered across campaign rounds. LLM = \texttt{GPT-5.6-Luna}.}
  \label{fig:SI-recall-per-material-luna}
\end{figure}

\begin{figure}[H]
  \centering
  \includegraphics[width=1.0\linewidth]{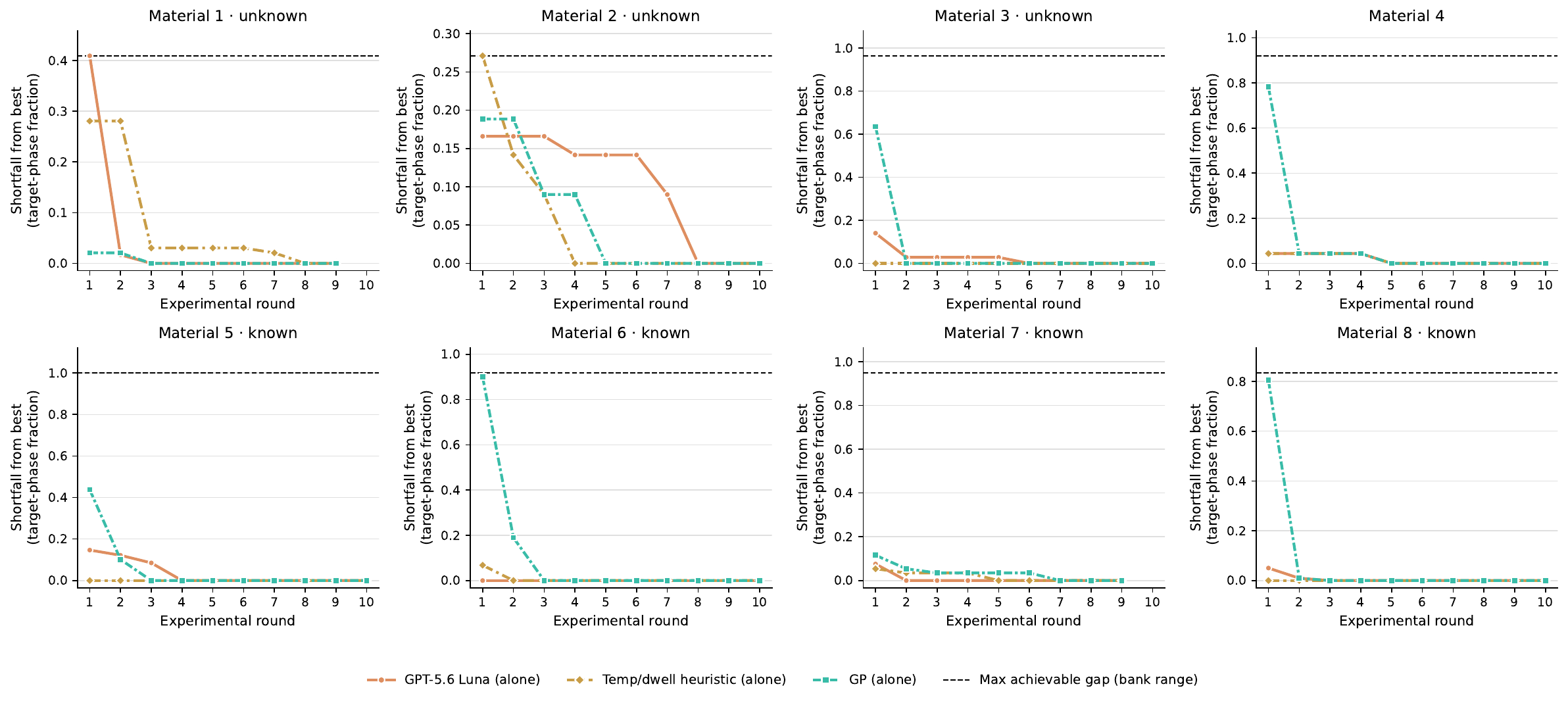}%
    \caption{Per-material simple regret across campaign rounds. LLM = \texttt{GPT-5.6-Luna}.}
  \label{fig:SI-regret-per-material-luna}
\end{figure}

%%%%%%%%%%%%%%%%%%%%%%%%%%%%%%%%%%%%%%%%%%%%%%%%%%%%%%%%%%%%
\end{document}